# Twitter mood predicts the stock market.

Johan Bollen[1,*], Huina Mao[1,*], Xiao-Jun Zeng[2].

*: authors made equal contributions.

*Abstract*—Behavioral economics tells us that emotions can profoundly affect individual behavior and decision-making. Does this also apply to societies at large, i.e. can societies experience mood states that affect their collective decision making? By extension is the public mood correlated or even predictive of economic indicators? Here we investigate whether measurements of collective mood states derived from large-scale Twitter feeds are correlated to the value of the Dow Jones Industrial Average (DJIA) over time. We analyze the text content of daily Twitter feeds by two mood tracking tools, namely OpinionFinder that measures positive vs. negative mood and Google-Profile of Mood States (GPOMS) that measures mood in terms of 6 dimensions (Calm, Alert, Sure, Vital, Kind, and Happy). We cross-validate the resulting mood time series by comparing their ability to detect the public's response to the presidential election and Thanksgiving day in 2008. A Granger causality analysis and a Self-Organizing Fuzzy Neural Network are then used to investigate the hypothesis that public mood states, as measured by the OpinionFinder and GPOMS mood time series, are predictive of changes in DJIA closing values. Our results indicate that the accuracy of DJIA predictions can be significantly improved by the inclusion of specific public mood dimensions but not others. We find an accuracy of 87.6% in predicting the daily up and down changes in the closing values of the DJIA and a reduction of the Mean Average Percentage Error by more than 6%.

## I. INTRODUCTION

STOCK market prediction has attracted much attention from academia as well as business. But can the stock market really be predicted? Early research on stock market prediction [1], [2], [3] was based on random walk theory and the Efficient Market Hypothesis (EMH) [4]. According to the EMH stock market prices are largely driven by *new* information, i.e. news, rather than present and past prices. Since news is unpredictable, stock market prices will follow a random walk pattern and cannot be predicted with more than 50 percent accuracy [5].

There are two problems with EMH. First, numerous studies show that stock market prices do not follow a random walk and can indeed to some degree be predicted [5], [6], [7], [8] thereby calling into question EMH's basic assumptions. Second, recent research suggests that news may be unpredictable but that very early indicators can be extracted from online social media (blogs, Twitter feeds, etc) to predict changes in various economic and commercial indicators. This may conceivably also be the case for the stock market. For example, [11] shows how online chat activity predicts book sales. [12] uses assessments of blog sentiment to predict movie sales. [15] predict future product sales using a Probabilistic Latent Semantic Analysis (PLSA) model to extract indicators of sentiment from blogs. In addition, Google search queries have been shown to provide early indicators of disease infection rates and consumer spending [14]. [9] investigates the relations between breaking financial news and stock price changes. Most recently [13] provide a ground-breaking demonstration of how public sentiment related to movies, as expressed on Twitter, can actually predict box office receipts.

Although news most certainly influences stock market prices, public mood states or sentiment may play an equally important role. We know from psychological research that emotions, in addition to information, play an significant role in human decision-making [16], [18], [39]. Behavioral finance has provided further proof that financial decisions are significantly driven by emotion and mood [19]. It is therefore reasonable to assume that the public mood and sentiment can drive stock market values as much as news. This is supported by recent research by [10] who extract an indicator of public anxiety from LiveJournal posts and investigate whether its variations can predict S&P500 values.

However, if it is our goal to study how public mood influences the stock markets, we need reliable, scalable and early assessments of the public mood at a time-scale and resolution appropriate for practical stock market prediction. Large surveys of public mood over representative samples of the population are generally expensive and time-consuming to conduct, cf. Gallup's opinion polls and various consumer and well-being indices. Some have therefore proposed indirect assessment of public mood or sentiment from the results of soccer games [20] and from weather conditions [21]. The accuracy of these methods is however limited by the low degree to which the chosen indicators are expected to be correlated with public mood.

Over the past 5 years significant progress has been made in sentiment tracking techniques that extract indicators of public mood directly from social media content such as blog content [10], [12], [15], [17] and in particular large-scale Twitter feeds [22]. Although each so-called *tweet*, i.e. an individual user post, is limited to only 140 characters, the aggregate of millions of tweets submitted to Twitter at any given time may provide an accurate representation of public mood and sentiment. This has led to the development of real-time sentiment-tracking indicators such as [17] and "Pulse of Nation"[1].

In this paper we investigate whether public sentiment, as expressed in large-scale collections of daily Twitter posts, can be used to predict the stock market. We use two tools to measure variations in the public mood from tweets submitted to the Twitter service from February 28, 2008 to December 19, 2008. The first tool, OpinionFinder, analyses the text content of tweets submitted on a given day to provide a positive vs. negative daily time series of public mood. The second tool, GPOMS, similarly analyses the text content of tweets to generate a six-dimensional daily time series of public mood to provide a more detailed view of changes in public along a variety of different mood dimensions. The resulting public mood time series are correlated to the Dow Jones Industrial Average (DJIA) to assess their ability to predict changes in the DJIA over time. Our results indicate that the prediction accuracy of standard stock market prediction models is significantly improved when certain mood dimensions are included, but not others. In particular variations along the public mood dimensions of Calm and Happiness as measured by GPOMS seem to have a predictive effect, but not general happiness as measured by the OpinionFinder tool.

---

[1]http://www.ccs.neu.edu/home/amislove/twittermood/

## II. RESULTS

### A. Data and methods overview

We obtained a collection of public tweets that was recorded from February 28 to December 19th, 2008 (9,853,498 tweets posted by approximately 2.7M users). For each tweet these records provide a tweet identifier, the date-time of the submission (GMT+0), its submission type, and the text content of the Tweet which is by design limited to 140 characters. After removal of stop-words and punctuation, we group all tweets that were submitted on the same date. We only take into account tweets that contain explicit statements of their author's mood states, i.e. those that match the expressions "i feel","i am feeling","i'm feeling","i dont feel", "I'm", "Im", "I am", and "makes me". In order to avoid spam messages and other information-oriented tweets, we also filter out tweets that match the regular expressions "http:" or "www."

As shown in Fig. 1 we then proceed in three phases. In the first phase, we subject the collections of daily tweets to 2 mood assessment tools: (1) OpinionFinder which measures positive vs. negative mood from text content, and (2) GPOMS which measures 6 different mood dimensions from text content. This results in a total of 7 public mood time series, one generated by OpinionFinder and six generated by GPOMS, each representing a potentially different aspect of the public's mood on a given day. In addition, we extract a time series of daily DJIA closing-values from Yahoo! Finance. In the second phase, we investigate the hypothesis that public mood as measured by GPOMS and OpinionFinder is predictive of future DJIA values. We use a Granger causality analysis in which we correlate DJIA values to GPOMs and OF values of the past $n$ days. In the third phase, we deploy a Self-Organizing Fuzzy Neural Network model to test the hypothesis that the prediction accuracy of DJIA prediction models can be improved by including measurements of public mood. We are not interested in proposing an optimal DJIA prediction model, but to assess the effects of including public mood information on the accuracy of a "baseline" prediction model.

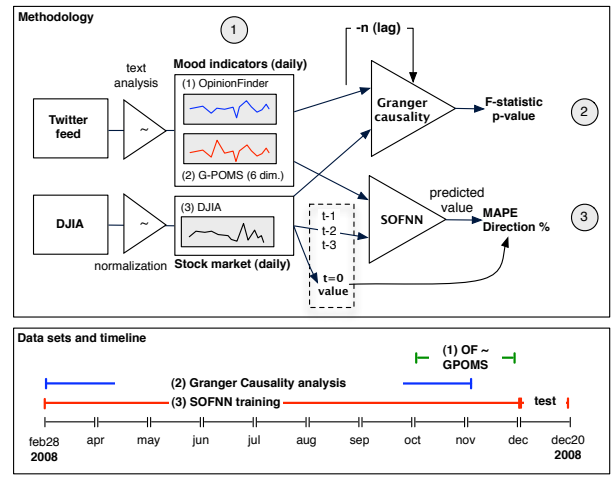

Fig. 1. Diagram outlining 3 phases of methodology and corresponding data sets: (1) creation and validation of OpinionFinder and GPOMS public mood time series from October 2008 to December 2008 (Presidential Election and Thanksgiving), (2) use of Granger causality analysis to determine correlation between DJIA, OpinionFinder and GPOMS public mood from August 2008 to December 2008, and (3) training of a Self-Organizing Fuzzy Neural Network to predict DJIA values on the basis of various combinations of past DJIA values and OF and GPOMS public mood data from March 2008 to December 2008.

### B. Generating public mood time series: OpinionFinder and GPOMS

OpinionFinder (OF)[2] is a publicly available software package for sentiment analysis that can be applied to determine sentence-level subjectivity [25], i.e. to identify the emotional polarity (positive or negative) of sentences. It has been successfully used to analyze the emotional content of large collections of tweets [26] by using the OF lexicon to determine the ratio of positive versus negative tweets on a given day. The resulting time series were shown to correlate with the Consumer Confidence Index from Gallup[3] and the Reuters/University of Michigan Surveys of Consumers[4] over a given period of time. We adopt OF's subjective lexicon that has been established upon previous work [37], [38], [24]. We select positive and negative words that are marked as either "weak" and "strong" from the OF sentiment lexicon resulting in a list of 2718 positive and 4912 negative words. For each tweet we determine whether it contains any number of negative and positive terms from the OF lexicon. For each occurrence we increase the score of either negative or positive tweets by 1 and calculate the ratio of positive vs. negative messages for the tweets posted on the same day $t$.

Like many sentiment analysis tools OF adheres to a unidimensional model of mood, making binary distinctions between positive and negative sentiment [23]. This may however ignore the rich, multi-dimensional structure of human mood. To capture additional dimensions of public mood we created a second mood analysis tools, labeled GPOMS, that can measure human mood states in terms of 6 different mood dimensions, namely *Calm*, *Alert*, *Sure*, *Vital*, *Kind* and *Happy*. GPOMS' mood dimensions and lexicon are derived from an existing and well-vetted psychometric instrument, namely the Profile of Mood States (POMS-bi)[32], [33]. To make it applicable to Twitter mood analysis we expanded the original 72 terms of the POMS questionnaire to a lexicon of 964 associated terms by analyzing word co-occurrences in a collection of 2.5 billion 4- and 5-grams[5] computed by Google in 2006 from approximately 1 trillion word tokens observed in publicly accessible Webpages [35], [36]. The enlarged lexicon of 964 terms thus allows GPOMS to capture a much wider variety of naturally occurring mood terms in Tweets and map them to their respective POMS mood dimensions. We match the terms used in each tweet against this lexicon. Each tweet term that matches an n-gram term is mapped back to its original POMS terms (in accordance with its co-occurence weight) and via the POMS scoring table to its respective POMS dimension. The score of each POMS mood dimension is thus determined as the weighted sum of the co-occurence weights of each tweet term that matched the GPOMS lexicon. All data sets and methods are available on our project web site[6].

---

[2]http://www.cs.pitt.edu/mpqa/opinionfinderrelease/
[3]http://www.gallup.com/poll/122840/Gallup-Daily-Economic-Indexes.aspx
[4]http://www.sca.isr.umich.edu/

To enable the comparison of OF and GPOMS time series we normalize them to z-scores on the basis of a local mean and standard deviation within a sliding window of $k$ days before and after the particular date. For example, the z-score of time series $X_t$, denoted $\mathbb{Z}_{X_t}$, is defined as:

$$\mathbb{Z}_{X_t} = \frac{X_t - \bar{x}(X_{t\pm k})}{\sigma(X_{t\pm k})} \quad (1)$$

where $\bar{x}(X_{t\pm k})$ and $\sigma(D_{t\pm k})$ represent the mean and standard deviation of the time series within the period $[t-k, t+k]$. This normalization causes all time series to fluctuate around a zero mean and be expressed on a scale of 1 standard deviation.

### C. Cross-validating OF and GPOMS time series against large socio-cultural events

We first validate the ability of OF and GPOMS to capture various aspects of public mood. To do so we apply them to tweets posted in a 3-month period from October 5, 2008 to December 5, 2008. This period was chosen specifically because it includes several socio-cultural events that may have had a unique, significant and complex effect on public mood namely the U.S presidential election (November 4, 2008) and Thanksgiving (November 27, 2008). The OF and GPOMS measurements can therefore be cross-validated against the expected emotional responses to these events. The resulting mood time series are shown in Fig. 2 and are expressed in z-scores as given by in Eq. 1.

Fig. 2 shows that the OF successfully identifies the public's emotional response to the Presidential election on November 4th and Thanksgiving on November 27th. In both cases OF marks a significant, but short-lived uptick in positive sentiment specific to those days.

The GPOMS results reveal a more differentiated public mood response to the events in the three-day period surrounding the election day (November 4, 2008). November 3, 2008 is

[5]n-grams are frequently occurring sequences of terms in text of length $n$, for example "we are the robots" could be a frequent 4-gram.
[6]http://terramood.informatics.indiana.edu/data

characterized by a significant drop in Calm indicating highly elevated levels of public anxiety. Election Day itself is characterized by a reversal of Calm scores indicating a significant reduction in public anxiety, in conjunction with a significant increases of Vital, Happy as well as Kind scores. The latter indicates a public that is energized, happy and friendly on election day. On November 5, these GPOMS dimensions continue to indicate positive mood levels, in particular high levels of Calm, Sure, Vital and Happy. After November 5, all mood dimensions gradually return to the baseline. The public mood response to Thanksgiving on November 27, 2008 provides a counterpart to the differentiated response to the Presidential election. On Thanksgiving day we find a spike in Happy values, indicating high levels of public happiness. However, no other mood dimensions are elevated on November 27. Furthermore, the spike in Happy values is limited to the one day, i.e. we find no significant mood response the day before or after Thanksgiving.

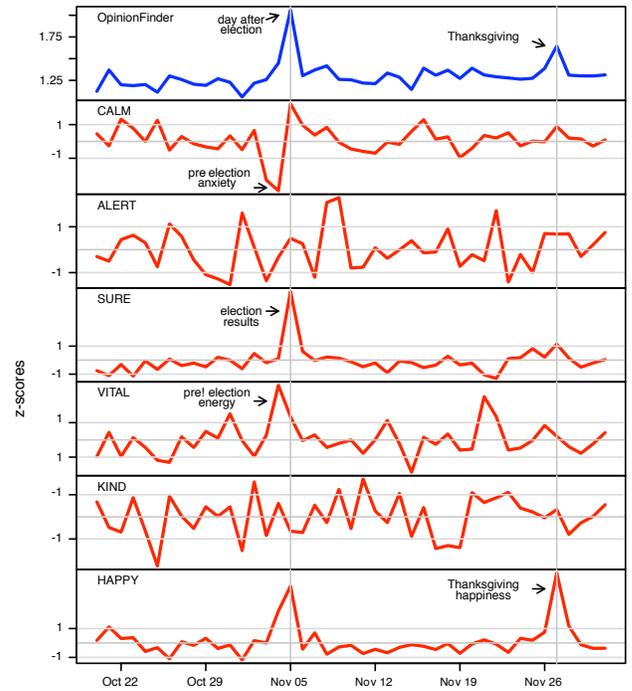

Fig. 2. Tracking public mood states from tweets posted between October 2008 to December 2008 shows public responses to presidential election and thanksgiving.

A visual comparison of Fig. 2 suggests that GPOMS' Happy dimension best approximates the mood trend provided by OpinionFinder. To quantitatively determine the relations between GPOMS's mood dimensions and the OF mood trends, we test the correlation between the trend obtained from OF lexicon and the six dimensions of GPOMS using multiple regression. The regression model is shown in Eq. 2.

$$Y_{OF} = \alpha + \sum_{i}^{n} \beta_i X_i + \epsilon_t \quad (2)$$

where $X_1$, $X_2$, $X_3$, $X_4$, $X_5$ and $X_6$ represent the mood time series obtained from the 6 GPOMS dimensions, respectively

TABLE I
MULTIPLE REGRESSION RESULTS FOR OPINIONFINDER VS. 6 GPOMS MOOD DIMENSIONS.

| Parameters | Coeff. | Std.Err. | t | p |
|---|---|---|---|---|
| Calm ($X_1$) | 1.731 | 1.348 | 1.284 | 0.20460 |
| Alert ($X_2$) | 0.199 | 2.319 | 0.086 | 0.932 |
| Sure ($X_3$) | 3.897 | 0.613 | 6.356 | 4.25e-08 ⋆⋆⋆ |
| Vital ($X_4$) | 1.763 | 0.595 | 2.965 | 0.004⋆⋆ |
| Kind ($X_5$) | 1.687 | 1.377 | 1.226 | 0.226 |
| Happy ($X_6$) | 2.770 | 0.578 | 4.790 | 1.30e-05 ⋆⋆ |
| Summary | Residual Std.Err | Adj.$R^2$ | $F_{6,55}$ | p |
| | 0.078 | 0.683 | 22.93 | 2.382e-13 |

(p-value < 0.001: ⋆⋆⋆, p-value < 0.05: ⋆⋆, p-value < 0.1: ⋆)

Calm, Alert, Sure, Vital, Kind and Happy.

The multiple linear regression results are provided in Table I (coefficient and p-values), and indicate that $Y_{OF}$ is significantly correlated with $X_3$ (Sure), $X_4$ (Vital) and $X_6$ (Happy), but not with $X_1$ (Calm), $X_2$ (Alert) and $X_5$ (Kind). We therefore conclude that certain GPOMS mood dimension partially overlap with the mood values provided by OpinionFinder, but not necessarily all mood dimensions that may be important in describing the various components of public mood e.g. the varied mood response to the Presidential election. The GPOMS thus provides a unique perspective on public mood states not captured by uni-dimensional tools such as OpinionFinder.

### D. Bivariate Granger Causality Analysis of Mood vs. DJIA prices

After establishing that our mood time series responds to significant socio-cultural events such as the Presidential election and Thanksgiving, we are concerned with the question whether other variations of the public's mood state correlate with changes in the stock market, in particular DJIA closing values. To answer this question, we apply the econometric technique of Granger causality analysis to the daily time series produced by GPOMS and OpinionFinder vs. the DJIA. Granger causality analysis rests on the assumption that if a variable $X$ causes $Y$ then changes in $X$ will systematically occur before changes in $Y$. We will thus find that the lagged values of $X$ will exhibit a statistically significant correlation with $Y$. Correlation however does not prove causation. We therefore use Granger causality analysis in a similar fashion to [10]; we are not testing actual causation but whether one time series has predictive information about the other or not[7].

Our DJIA time series, denoted $D_t$, is defined to reflect daily changes in stock market value, i.e. its values are the delta between day $t$ and day $t-1$: $D_t = DJIA_t - DJIA_{t-1}$. To test whether our mood time series predicts changes in stock market values we compare the variance explained by two linear models as shown in Eq. 3 and Eq. 4. The first model ($L_1$) uses only $n$ lagged values of $D_t$, i.e. ($D_{t-1}, \cdots, D_{t-n}$) for prediction, while the second model $L_2$ uses the $n$ lagged values of both $D_t$ and the GPOMS plus the OpinionFinder mood time series denoted $X_{t-1}, \cdots, X_{t-n}$.

[7][10] uses only one mood index, namely Anxiety, but we investigate the relation between DJIA values and all Twitter mood dimensions measured by GPOMS and OpinionFinder

We perform the Granger causality analysis according to model $L_1$ and $L_2$ shown in Eq. 3 and 4 for the period of time between February 28 to November 3, 2008 to exclude the exceptional public mood response to the Presidential Election and Thanksgiving from the comparison. GPOMS and OpinionFinder time series were produced for 342,255 tweets in that period, and the daily Dow Jones Industrial Average (DJIA) was retrieved from Yahoo! Finance for each day[8].

$$L_1 : D_t = \alpha + \sum_{i=1}^{n} \beta_i D_{t-i} + \epsilon_t \quad (3)$$

$$L_2 : D_t = \alpha + \sum_{i=1}^{n} \beta_i D_{t-i} + \sum_{i=1}^{n} \gamma_i X_{t-i} + \epsilon_t \quad (4)$$

Based on the results of our Granger causality (shown in Table II), we can reject the null hypothesis that the mood time series do not predict DJIA values, i.e. $\beta_{\{1,2,\cdots,n\}} \neq 0$ with a high level of confidence. However, this result only applies to 1 GPOMS mood dimension. We observe that $X_1$ (i.e. Calm) has the highest Granger causality relation with DJIA for lags ranging from 2 to 6 days (p-values < 0.05). The other four mood dimensions of GPOMS do not have significant causal relations with changes in the stock market, and neither does the OpinionFinder time series.

To visualize the correlation between $X_1$ and the DJIA in more detail, we plot both time series in Fig. 3. To maintain the same scale, we convert the DJIA delta values $D_t$ and mood index value $X_t$ to z-scores as shown in Eq. 1.

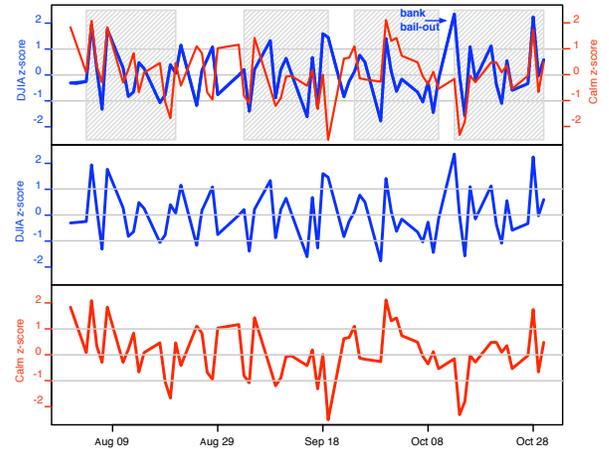

Fig. 3. A panel of three graphs. The top graph shows the overlap of the day-to-day difference of DJIA values (blue: $\mathbb{Z}_{D_t}$) with the GPOMS' Calm time series (red: $\mathbb{Z}_{X_t}$) that has been lagged by 3 days. Where the two graphs overlap the Calm time series predict changes in the DJIA closing values that occur 3 days later. Areas of significant congruence are marked by gray areas. The middle and bottom graphs show the separate DJIA and GPOMS' Calm time series.

As can be seen in Fig. 3 both time series frequently overlap or point in the same direction. Changes in past values of Calm ($t-3$) predicts a similar rise or fall in DJIA values ($t =$

[8]Our DJIA time series has no values for weekends and holidays because trading is suspended during those days. We do not linearly extropolate to fill the gaps. This results in a time series of 64 days.

TABLE II
STATISTICAL SIGNIFICANCE (P-VALUES) OF BIVARIATE GRANGER-CAUSALITY CORRELATION BETWEEN MOODS AND DJIA IN PERIOD FEBRUARY 28, 2008 TO NOVEMBER 3, 2008.

| Lag | OF | Calm | Alert | Sure | Vital | Kind | Happy |
|---|---|---|---|---|---|---|---|
| 1 day | **0.085*** | 0.272 | 0.952 | 0.648 | 0.120 | 0.848 | 0.388 |
| 2 days | 0.268 | **0.013**** | 0.973 | 0.811 | 0.369 | 0.991 | 0.7061 |
| 3 days | 0.436 | **0.022**** | 0.981 | 0.349 | 0.418 | 0.991 | 0.723 |
| 4 days | 0.218 | **0.030**** | 0.998 | 0.415 | 0.475 | 0.989 | 0.750 |
| 5 days | 0.300 | **0.036**** | 0.989 | 0.544 | 0.553 | 0.996 | 0.173 |
| 6 days | 0.446 | **0.065*** | 0.996 | 0.691 | 0.682 | 0.994 | **0.081*** |
| 7 days | 0.620 | 0.157 | 0.999 | 0.381 | 0.713 | 0.999 | 0.150 |

(p-value $< 0.05$: $\star\star$, p-value $< 0.1$: $\star$)

0. The Calm mood dimension thus has predictive value with regards to the DJIA. In fact the p-value for this shorter period, i.e. August 1, 2008 to October 30 2008, is significantly lower (lag $n = 3$, $p = 0.009$) than that listed in Table II for the period February 28, 2008 to November 3, 2008.

The cases in which the $t - 3$ mood time series fails to track changes in the DJIA are nearly equally informative as where it doesn't. In particular we point to a significant deviation between the two graphs on October 13th where the DJIA surges by more than 3 standard deviations trough-to-peak. The Calm curve however remains relatively flat at that time after which it starts to again track changes in the DJIA again. This discrepancy may be the result of the the Federal Reserve's announcement on October 13th of a major bank bailout initiative which unexpectedly increase DJIA values that day. The deviation between Calm values and the DJIA on that day illustrates that unexpected news is not anticipated by the public mood yet remains a significant factor in modeling the stock market.

*E. Non-linear models for emotion-based stock prediction*

Our Granger causality analysis suggests a predictive relation between certain mood dimensions and DJIA. However, Granger causality analysis is based on linear regression whereas the relation between public mood and stock market values is almost certainly non-linear. To better address these non-linear effects and assess the contribution that public mood assessments can make in predictive models of DJIA values, we compare the performance of a Self-organizing Fuzzy Neural Network (SOFNN) model [30] that predicts DJIA values on the basis of two sets of inputs: (1) the past 3 days of DJIA values, and (2) the same combined with various permutations of our mood time series (explained below). Statistically significant performance differences will allow us to either confirm or reject the null hypothesis that public mood measurement do not improve predictive models of DJIA values.

We use a SOFNN as our prediction model since they have previously been used to decode nonlinear time series data which describe the characteristics of the stock market [28] and predict its values [29]. Our SOFNN in particular is a five-layer hybrid neural network with the ability to self-organize its own neurons in the learning process. A similar organization has been successfully used for electricial load forecasting in our previous work [31].

To predict the DJIA value on day $t$, the input attributes of our SOFNN include combinations of DJIA values and mood values of the past $n$ days. We choose $n = 3$ since the results shown in Table II indicate that past $n = 4$ the Granger causal relation between Calm and DJIA decreases significantly. All historical load values are linearly scaled to [0,1]. This procedure causes every input variable be treated with similar importance since they are processed within a uniform range.

SOFNN models require the tuning of a number of parameters that can influence the performance of the model. We maintain the same parameter values across our various input combinations to allow an unbiased comparison of model performance, namely $\delta = 0.04, \sigma = 0.01, k_{rmse} = 0.05, k_d(i), (i = 1, ..., r) = 0.1$ where $r$ is the dimension of input variables and $k_{rmse}$ is the expected training root mean squared error which is a predefined value.

To properly evaluate the SOFNN model's ability to predict daily DJIA prices, we extend the period under consideration to February 28, 2008 to December 19, 2008 for training and testing. February 28, 2008 to November 28, 2008 is chosen as the longest possible training period while Dec 1 to Dec 19, 2008 was chosen as the test period because it was characterized by stabilization of DJIA values after considerable volatility in previous months and the absence of any unusual or significant socio-cultural events. Fig. 4 shows that the Fall of 2008 is an unusual period for the DJIA due to a sudden dramatic decline of stock prices. This variability may in fact render stock market prediction more difficult than in other periods.

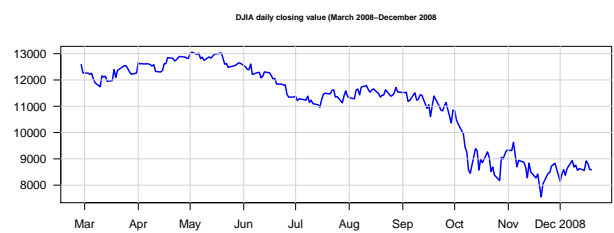

Fig. 4. Daily Dow Jones Industrial Average values between February 28, 2008 and December 19, 2008.

The Granger causality analysis indicates that only Calm (and to some degree Happy) is Granger-causative of DJIA values. However, the other mood dimensions could still contain predictive information of DJIA values when combined with other mood dimensions. For example, Happy may not be independently linearly related with DJIA, but it may nevertheless improve the SOFNN prediction accuracy when combined with Calm. To clarify these questions, we investigate seven permutations of input variables to the SOFNN model, the first of which, denoted $I_0$, represents a naive, baseline model that has been trained to predict DJIA values at time $t$ from the historical values at time $\{t-1, t-2, t-3\}$:

$$\begin{aligned}
I_0 &= \{\text{DJIA}_{t-3,2,1}\} \\
I_1 &= \{\text{DJIA}_{t-3,2,1}, X_{1,t-3,2,1}\} \\
I_{1,2} &= \{\text{DJIA}_{t-3,2,1}, X_{1,t-3,2,1}, X_{2,t-3,2,1}\} \\
I_{1,3} &= \{\text{DJIA}_{t-3,2,1}, X_{1,t-3,2,1}, X_{3,t-3,2,1}\} \\
&\ldots
\end{aligned}$$

$\text{DJIA}_{t-3,2,1}$ represents the DJIA values and $X_{1,t-3,2,1}$ represents the values of the GPOMS mood dimension 1, at time $t-3$, $t-2$, and $t-1$. According to the same notation $I_{1,3}, I_{1,4}, I_{1,5}, I_{1,6}$ represent a combination of historical DJIA with mood dimensions 3, 4, 5 and 6 at time $t-3$, $t-2$, and $t-1$. For example, $I_{1,6}$ represents a set of inputs that includes the DJIA values $t-3$, $t-2$, and $t-1$, and mood dimensions 1 and 6 at the same times.

In order to compare the performance of the GPOMS mood data with the positive/negative sentiment values given by OpinionFinder, we additionally define the input combination:

$$I_{\text{OF}} = \{\text{DJIA}_{t-3,2,1}, X_{\text{OF},t-3,2,1}\}$$

Forecasting accuracy is measured in terms of the average Mean Absolute Percentage Error (MAPE) and the direction accuracy (up or down) during the test period (December 1 to December 19, 2008). The prediction results are shown in Table III.

We can draw several conclusions from these results. First, adding positive/negative sentiment obtained from OF ($I_{\text{OF}}$) has no effect on prediction accuracy compared to using only historical DJIA values($I_0$). This confirms the results of our Granger causality analysis.

Second, adding Calm, i.e. input $I_1$, we find the highest prediction accuracy. Compared to $I_0$ and all other input combinations, adding input $I_1$ leads to significant improvements in MAPE values (1.83% vs. the maximum of 2.13% and 1.95% for $I_{\text{OF}}$) and direction accuracy (86.7% compared to 73.3% for $I_{\text{OF}}$ and 46.7% for $I_{1,3}$). Thirdly, $I_{1,3}$ and $I_{1,4}$ actually reduce prediction accuracy significantly both in terms of MAPE and direction %, most likely because $X_3$ (Sure) and $X_4$ (Vital) do not contain information that is useful for prediction of DJIA values causing prediction accuracy to regress to chance levels. It is notable that $I_{1,6}$, i.e. a combination of $X_6$ and $X_1$ does significantly reduce average MAPE, and provides good direction accuracy (80%). This is surprising since $X_6$ (i.e. Happy) does not have a good Granger causality relation with DJIA at a lag of $n = 3$ days (see Table II, $p = 0.723$). However in combination with Calm, it produces a more accurate SOFNN prediction (MAPE=1.79%) and direction accuracy (80%).

To assess the statistical significance of the SOFNN achieving the above mentioned accuracy of 87.6% in predicting the up and down movement of the DJIA we calculate the odds of this result occurring by chance. The binomial distribution indicates that the probability of achieving exactly 87.6% correct guesses over 15 trials (20 days minus weekends) with a 50% chance of success on each single trial equals 0.32%. Taken over the entire length of our data set (February 28 to December 20, excluding weekends) we find approximately 10.9 of such 20 day periods. The odds that the mentioned probability would hold by chance for a random period of 20 days within that period is then estimated to be $1-(1-0.0032)^{10.9} = 0.0343$ or 3.4%. The SOFNN direction accuracy is thus most likely not the result of chance nor our selecting a specifically favorable test period.

In addition, we test the linear effect of both Calm($X_1$) and Happy ($X_6$) on DJIA, with a nested $F-test$ between the full model $F$ and reduced model $R$ shown as follows:

$$\mathbf{F}: \quad D_t = \alpha + \sum_{i=1}^{n} \beta_i D_{t-i} + \sum_{i=1}^{3} \gamma_i X_{1,t-i} + \sum_{i=1}^{3} \gamma_i X_{6,t-i} + \epsilon_t$$

$$\mathbf{R}: \quad D_t = \alpha + \sum_{i=1}^{n} \beta_i D_{t-i} + \sum_{i=1}^{3} \gamma_i X_{1,t-i} + \epsilon_t$$

We find a p-value of 0.66 and an F-statistic of 0.53 indicating that a linear combination of $X_1$ and $X_6$ produces worse results than $X_1$ alone. Since the SOFNN prediction is more accurate when using a combination of $X_1$ and $X_6$, we conclude that this confirms a nonlinear relation among the different dimensions of moods.

## III. DISCUSSION

In this paper, we investigate whether public mood as measured from large-scale collection of tweets posted on twitter.com is correlated or even predictive of DJIA values. Our results show that changes in the public mood state can indeed be tracked from the content of large-scale Twitter feeds by means of rather simple text processing techniques and that such changes respond to a variety of socio-cultural drivers in a highly differentiated manner. Among the 7 observed mood dimensions only some are Granger causative of the DJIA; changes of the public mood along these mood dimensions match shifts in the DJIA values that occur 3 to 4 days later. Surprisingly we do not observe this effect for OpinionFinder's assessment of public mood states in terms of positive vs. negative mood but rather for the GPOMS dimension labeled "Calm". The calmness of the public (measured by GPOMS) is thus predictive of the DJIA rather than general levels of positive sentiment as measured by OpinionFinder. A Self-Organizing Fuzzy Neural Network trained on the basis of past DJIA values and our public mood time series furthermore demonstrated the ability of the latter to significantly improve the accuracy of even the most basic models to predict DJIA closing values. Given the performance increase for a relatively basic model such as the SOFNN we are hopeful to find equal or better improvements for more sophisticated market models that may in fact include other information derived from news sources, and a variety of relevant economic indicators. These results have implications for existing sentiment tracking tools as well as surveys of "self-reported subjective well-being" in which individuals evaluate the extent to which they experience positive and negative affect, happiness, or satisfaction with life [40]. Such surveys are relatively expensive and time-consuming, and may nevertheless not allow the measurement of public mood along mood dimensions that are relevant to assess particular socio-economic indicators. Public mood analysis from Twitter feeds on the other hand offers an automatic, fast, free and large-scale addition to this toolkit that may in addition be optimized to measure a variety of dimensions of the public mood state.

TABLE III
DJIA DAILY PREDICTION USING SOFNN

| Evaluation | $I_{OF}$ | $I_0$ | $I_1$ | $I_{1,2}$ | $I_{1,3}$ | $I_{1,4}$ | $I_{1,5}$ | $I_{1,6}$ |
|---|---|---|---|---|---|---|---|---|
| MAPE (%) | 1.95 | 1.94 | 1.83 | 2.03 | 2.13 | 2.05 | 1.85 | **1.79**★ |
| Direction (%) | 73.3 | 73.3 | **86.7**★ | 60.0 | 46.7 | 60.0 | 73.3 | 80.0 |

Our analysis does not acknowledge a number of important factors that will form the basis of future research. First, we note that our analysis is not designed to be limited to any particular geographical location nor subset of the world's population. This approach may be appropriate since the US stock markets are affected by individuals worldwide, but for the particular period under observation Twitter.com users were *de facto* predominantly English speaking and located in the US. As Twitter.com's user base becomes increasingly international and the use of smartphones equipped with geo-location increases, future analysis will have to factor in location and language to avoid geographical and cultural sampling errors. Second, although we have cross-validated the results of 2 different tools to assess public mood states, we have no knowledge of the "ground truth" for public mood states nor in fact for the particular subsample of the population represented by the community of Twitter.com users. This problem can only be addressed by increased research into direct assessments of public mood states vs. those derived from online communities such as Twitter. Third, these results are strongly indicative of a predictive correlation between measurements of the public mood states from Twitter feeds, but offer no information on the causative mechanisms that may connect public mood states with DJIA values in this manner. One could speculate that the general public is presently as strongly invested in the DJIA as financial experts, and that therefore their mood states will directly affect their investment decisions and thus stock market values, but this too remains an area of future research.

ACKNOWLEDGMENT

This research was supported by NSF Grant BCS #1032101. We thank David Crandell and Michael Nelson for their helpful comments on earlier versions of this manuscript. We are also grateful to Daniel Kahneman for his early comments on our work.

REFERENCES

[1] Fama, Eugene F, e. a. (1969) *International Economic Review* **10**, 1–21.
[2] Fama, E. F. (1991) *Journal of Finance* **46**, 1575–617.
[3] H.Cootner, P. (1964) *The random character of stock market prices*. (MIT).
[4] Fama, E. F. (1965) *The Journal of Business* **38**, 34–105.
[5] Qian, Bo, Rasheed, & Khaled. (2007) *Applied Intelligence* **26**, 25–33.
[6] Gallagher, L. A & Taylor, M. P. (2002) *Southern Economic Journal* **69**, 345–362.
[7] Kavussanos, M & Dockery, E. (2001) *Applied Financial Economics* **11**, 573–79.
[8] Butler, K. C & Malaikah, S. J. (1992) *Journal of Banking and Finance* **16**, 197–210.
[9] Schumaker, R. P & Chen, H. (2009) *ACM Trans. Inf. Syst.* **27**, 12:1–12:19.
[10] Gilbert, E & Karahalios, K. (2010) Widespread worry and the stock market.
[11] Gruhl, D, Guha, R, Kumar, R, Novak, J, & Tomkins, A. (2005) *The predictive power of online chatter*. (ACM, New York, NY, USA), pp. 78–87.
[12] Mishne, G & Glance, N. (2006) *Predicting Movie Sales from Blogger Sentiment*. AAAI 2006 Spring Symposium on Computational Approaches to Analysing Weblogs
[13] S. Asur and B. A. Huberman 2010 *Predicting the Future with Social Media* arXiv:1003.5699v1
[14] Choi, H & Varian, H. (2009) Predicting the present with google trends., (Google), Technical report.
[15] Liu, Y, Huang, X, An, A, & Yu, X. (2007) *ARSA: a sentiment-aware model for predicting sales performance using blogs*. (ACM, New York, NY, USA), pp. 607–614.
[16] Dolan, R. J. (2002) *Science* **298**, 1191–1194.
[17] Dodds, Peter. (2009) *Journal of Happiness* **July**, doi: 10.1007/s10902-009-9150-9
[18] Damasio, A. R. (1994) *Descartes' error : emotion, reason, and the human brain*. (Putnam), pp. xix, 312 p.+.
[19] Nofsinger, J. (2005) *Journal of Behaviour Finance.* **6**, 144–160.
[20] Edmans, A, Garca, D, & Norli, . (2007) *Journal of Finance* **62**, 1967–1998.
[21] Hirshleifer, D & Shumway, T. (2003) *Journal of Finance* **58**, 1009–1032.
[22] Pak, A & Paroubek, P. (2010) *Twitter as a Corpus for Sentiment Analysis and Opinion Mining*. (European Language Resources Association (ELRA), Valletta, Malta).
[23] Pang, B & Lee, L. (2008) *Foundations and Trends in Information Retrieval* **2**, 1–135.
[24] Wilson, T, Wiebe, J, & Hoffmann, P. (2005) *Recognizing Contextual Polarity in Phrase-Level Sentiment Analysis*. (Vancouver, CA).
[25] Wilson, T, Hoffmann, P, Somasundaran, S, Kessler, J, Wiebe, J, Choi, Y, Cardie, C, Riloff, E, & Patwardhan, S. (2005) *OpinionFinder: A system for subjectivity analysis*. pp. 34–35.
[26] O'Connor, B, Balasubramanyan, R, Routledge, B. R, & Smith, N. A. (2010) *From Tweets to Polls: Linking Text Sentiment to Public Opinion Time Series*.
[27] Diener E, Diener M & Diener D (2009) *Factors Predicting the Subjective Well-Being of Nations*. Social Indicators Research Series 38:43-70
[28] Lapedes, A & Farber, R. (1987) Nonlinear signal processing using neural network: Prediction and system modeling, (Los Alamos National Lab Technical Report), Technical report.
[29] Zhu, X, Wang, H, Xu, L, & Li, H. (2008) *Expert Syst. Appl.* **34**, 3043–3054.
[30] Leng, G, Prasad, G, & McGinnity, T. M. (2004) *Neural Netw.* **17**, 1477–1493.
[31] Mao, H, Zeng, X.-J, Leng, G, Zhai, Y, & Keane, A. J. (2009) *IEEE Transaction on Power System.* **24**, 1080–1090.
[32] Norcross, J. C, Guadagnoli, E, & Prochaska, J. O. (2006) *Journal of Clinical Psychology* **40**, 1270 – 1277.
[33] McNair, D, Heuchert, J. P, & Shilony, E. (2003) *Profile of mood states. Bibliography 1964–2002*. (Multi-Health Systems).
[34] Pepe, A & Bollen, J. (2008) *Between conjecture and memento: shaping a collective emotional perception of the future*.

[35] Brants, T & Franz, A. (2006) Web 1t 5-gram version 1, (Linguistic Data Consortium, Philadelphia), Technical report.
[36] Bergsma, S, Lin, D & Goebel, R (2009) *IJCAI'09: Proceedings of the 21st international jont conference on Artifical intelligence, San Francisco, CA.*, 1507–1512.
[37] Riloff, E, Wiebe, J, & Wilson, T. (2003) *Learning subjective nouns using extraction pattern bootstrapping*. (Association for Computational Linguistics, Morristown, NJ, USA), 25–32.
[38] Riloff, E & Wiebe, J. (2003) *Learning extraction patterns for subjective expressions*. (Association for Computational Linguistics, Morristown, NJ, USA), pp. 105–112.
[39] Kahneman, D & Tversky, Amos (1979) *Prospect Theory: An Analysis of Decision under Risk*. (Econometrica), pp. 263–291.
[40] Frey, B. S. (2008) *Happiness: A Revolution in Economics*. (The MIT Press).

**Johan Bollen** School of Informatics and Computing, Indiana Unviersity-Bloomington, United States Email:jbollen@indiana.edu

**Huina Mao** School of Informatics and Computing, Indiana Unviersity-Bloomington, United States Email: huinmao@indiana.edu

**Xiao-Jun Zeng** School of Computer Science, The University of Manchester, United Kingdom Email: x.zeng@manchester.ac.uk